\documentclass[11pt,aps,pra,showkeys]{revtex4-1}

\usepackage[breaklinks]{hyperref}
\usepackage[utf8]{inputenc}
\usepackage[OT2,T1]{fontenc}
\usepackage{graphicx}
\usepackage{color}
\usepackage{amsfonts,amsthm}
\usepackage{bm,bbm}
\usepackage{amsmath,amssymb,amsfonts} 

\newcommand{\be}{\begin{equation}}
\newcommand{\ee}{\end{equation}}
\newcommand{\ba}{\begin{eqnarray}}
\newcommand{\ea}{\end{eqnarray}}
\def\bs{\begin{subequations}}
\def\es{\end{subequations}}
\def\a{\alpha}

\def\k{\kappa}
\def\e{\epsilon}

\def\s{\sigma}

\def\cP{\mathcal{P}}

\def\ds{d_{\rm S}}

\newcommand{\Eq}[1]{(\ref{#1})}
\def\com{\color{magenta}}
\def\cob{\color{blue}}
\newcommand{\oarX}[1]{\href{http://arxiv.org/abs/#1}{{\ttfamily\com arXiv:#1}}}
\newcommand{\arX}[1]{\href{http://arxiv.org/abs/#1}{{\ttfamily\com arXiv:#1}}}
\newcommand{\doin}[6]{\href{http://dx.doi.org/#1}{{\cob #2 #3 {\bf #4}, #5 (#6)}}}
\newcommand{\doinn}[5]{\href{http://dx.doi.org/#1}{{\cob #2 {\bf #3}, #4 (#5)}}}
\newcommand{\doij}[5]{\href{http://dx.doi.org/#1}{{\cob #2 #3 (#5) #4}}}

\newcommand{\procm}[6]{in \emph{#1}, ed.\ by #2 (#3, #4, #5, #6)}
\newcommand{\procsinm}[5]{in \emph{#1}, ed.\ by #2 (#3, #4, #5)}
\newcommand{\tia}[1]{\textit{#1},}

\def\rme{\text{e}}
\def\rmd{\text{d}}

\begin{document}

\title{Entanglement entropy, scale-dependent dimensions and the origin of gravity\footnote{This essay received an Honorable Mention in the 2017 Essay Competition of the Gravity Research Foundation.}}

\date{\today}

\author{Michele Arzano}
\email{michele.arzano@roma1.infn.it}
\affiliation{Dipartimento di Fisica and INFN, ``Sapienza'' University of Rome, P.le A.\ Moro 2, 00185 Roma, Italy}

\author{Gianluca Calcagni}
\email{g.calcagni@csic.es (corresponding author)}
\affiliation{Instituto de Estructura de la Materia, CSIC, Serrano 121, 28006 Madrid, Spain}

\begin{abstract}
We argue that the requirement of a finite entanglement entropy of quantum degrees of freedom across a boundary surface is closely related to the phenomenon of running spectral dimension, universal in approaches to quantum gravity. If quantum geometry hinders diffusion, for instance when its structure at some given scale is discrete or too rough, then the spectral dimension of spacetime vanishes at that scale and the entropy density blows up. A finite entanglement entropy is a key ingredient in deriving Einstein gravity in a semi-classical regime of a quantum-gravitational theory and, thus, our arguments strengthen the role of running dimensionality as an imprint of quantum geometry with potentially observable consequences.
\end{abstract}


\date{March 31, 2017}



\centerline{\doin{10.1142/S0218271817430301}{Int.\ J.\ Mod.\ Phys.}{D}{26}{1743030}{2017}}

\maketitle

The connection between gravity and entropy, two seemingly disparate aspects of physics, was first unveiled in the seminal proposal by Bekenstein that black holes carry an entropy proportional to their area in Planck-area units \cite{BH1,BH2}. Since then, it has been a widely accepted view that such relationship holds one of the main keys to a deeper understanding of the quantum properties of spacetime. A variety of derivations of the entropy-area law have been proposed from drastically different approaches to quantum gravity (e.g., \cite{StVa,ABCK,OrPS}) strongly supporting the universality of the relation  and corroborating its role as one of the cornerstones of any attempt to solve the quandaries emerging in the interplay between gravity and the quantum. 

Recent years have witnessed an inversion of this trend. Rather than deriving the area-entropy relation from quantum gravity, the focus has shifted to understand gravity as a consequence of the quantum entropic properties of spacetime (see \cite{CHRR,CR} and references therein). At the core of these proposals is the assumption that, to any local horizon, one can associate an entropy proportional to its area. A natural candidate for this horizon entropy is the \emph{entanglement entropy} arising from quantum correlations between degrees of freedom of a quantum field across the horizon. 
 The problem is that, as it is well known, the microscopic structure of any local quantum field theory naturally predicts a divergent entanglement entropy density across a boundary. The main view is that quantization of geometry effectively cuts off the correlations rendering the entropy finite or, more suggestively, that a finite entanglement entropy density implies the existence of Einstein gravity as a semi-classical limit of a quantum theory. From this perspective, it becomes crucial to understand how a finite entanglement entropy can be realized in field theories which incorporate putative quantum-gravity effects via discreteness or new ultraviolet (UV) features beyond the structures of local quantum field theory. Below, we provide a first significative contribution in this direction.

As a starting point, let us give a closer look at the divergent character of the entanglement entropy density. The trace over the field-theoretical degrees of freedom associated with a region of space leads to an entanglement entropy which can be evaluated using appropriate techniques in Euclidean time \cite{Casini:2009sr,Solodukhin:2011gn}. The result is proportional to the ``area'' of the boundary $\Sigma$ times a divergent term,
\be\label{edens0}
S \sim A(\Sigma) \rho_d:=A(\Sigma) \int^{\infty}_{\epsilon^2} \frac{\rmd \s}{\s} \cP_{d} (\s)\,,
\ee
where the parameter $\s$ is the square of a fictitious diffusion time and $\e$ is a UV cut-off for small times or coarse resolutions. The function $\cP_{d}(\s)$ is the return probability for a diffusion process on the $d$-dimensional surface $\Sigma$ and it is given by the trace of the heat kernel associated with the two-point correlation function of a field theory living on the boundary \cite{Frolov:1998ea}. The integral $\rho_d$ in \eqref{edens0} can be interpreted as the entanglement entropy density whose divergent structure is readily understood: the short-distance behaviour of the correlation function $G(x,y)$ is related to the return probability $\cP_4(\s)\propto \int \rmd^4p\, \rme^{-\s C(p)}$, where $C(p)$ is the Fourier transform of the wave operator of the theory, by
\be\nonumber
\lim_{x\rightarrow y}G(x,y)\propto \int_0^{\infty} \rmd \s\, \cP_4(\s)\,.
\ee
Such limit usually diverges as the inverse power of the Euclidean distance between $x$ and $y$. This suggests that theories with a UV cut-off (introduced, e.g., through a modified inverse momentum space propagator $C(p)$) regularizing the short-distance behaviour of the two-point function should lead to a finite entanglement entropy. This naive expectation, however, is not met, since $\cP_d(\s)$ is always divergent in $\s=0$ due to the integration over a non-compact momentum space, no matter how regular the UV-behaviour of the propagator. This rules out all field theories with modified dispersion relations (which either break or preserve Lorentz invariance) as candidate models for a finite entanglement entropy density \cite{Nesterov:2010jh}.

This observation suggests that theories with finite $\cP_d(0)$ could realize our hope. A natural place to look for such theories is within models with {\it compact} boundary momentum space determined by a non-trivial four-momentum space geometry \cite{KowalskiGlikman:2004tz,Girelli:2009yz,Arzano:2010jw,Alesci:2011cg,Arzano:2014jfa,Amelino-Camelia:2013gna}. Also in this case, however, one concludes that, despite exhibiting a finite return probability in the vanishing diffusion time limit, the associated entanglement entropy density is still divergent \cite{ArCa3}. To appreciate this, we must turn to the analysis of the spectral zeta function of the return probability on the $d$-dimensional boundary and this will naturally lead to consider the notion of {\it spectral dimension}. The zeta function is defined as the Mellin transform of the return probability,
\be\label{zetad}
\zeta_d(s):=\zeta_{\cP_{d}}(s)=\frac{1}{\Gamma(s)}\int_0^{+\infty}\rmd \s\,\s^{s-1}\cP_{d}(\s)\,.
\ee
In the quantum gravity literature, the spectral dimension $\ds$ of spacetime (in particular, of the spatial boundary, which we denote by $\ds^{\rm b}$) is usually computed as the scaling of the return probability, $\cP_d(\s)\sim\s^{-\ds/2}$. In usual $D$-dimensional flat Minkowski space, the spectral dimension is constant and coincides with the topological dimension $D$. Over the past decade, evidence has been accumulating suggesting that, as one approaches the Planck scale, the dimensionality of space undergoes a change of value captured by the behaviour of the spectral dimension (e.g., \cite{tH93,Car09,revmu}). 

To establish a direct link between a running spectral dimension and entanglement entropy, here we take a less explored view on the former, more familiar to the field of fractal geometry \cite{Akk12}. From this perspective, the spectral dimension is nothing but the real part of the poles of the spectral zeta function. In scale-invariant geometries (fractals or ordinary space: constant $\ds$), one takes a small-$\s$ expansion and considers only the pole with largest real part ${\rm Re}(s)=\ds/2$. In multiscale geometries (multifractals or quantum gravity: running $\ds$), one can extend this discussion to an abstract boundary with $d=D-2$ topological dimensions and argue \cite{ArCa3} that the real parts of the poles of $\zeta_d(s)$ determine the non-vanishing plateaux of the spectral dimension $\ds^{\rm b}$ of the boundary. Moreover, if $\ds^{\rm b}=0$ at some point or plateau in dimensional flow, then $\zeta_d(0)$ is finite. These observations combine in a rather intriguing fashion that establishes a new view on the entanglement entropy. Noting from \Eq{edens0} and \Eq{zetad} that
\be\label{eefin}
\rho_d=\lim_{s\to 0} \Gamma(s)\,\zeta_d(s)\,,
\ee
we realize that \emph{it is impossible to have a finite entanglement entropy density when the spectral dimension $\ds^{\rm b}$ of the spatial boundary vanishes at some scale}. Thus, a necessary condition for a finite entropy density is that $\ds^{\rm b}\neq 0$ at all scales. Physically, a plateau $\ds^{\rm b}= 0$ implies a constant return probability in a given range of scales, meaning that a test particle does not diffuse on the boundary; this can happen if the boundary geometry is too irregular or disconnected at those scales. Therefore, a finite entropy can arise only in geometries sufficiently smooth, non-degenerate, or not pathologically rough (the irregularity of fractals is still acceptable; see below).

Let us illustrate this result with some examples. Field theories on a compact momentum space do not give rise to a finite $\rho_d$. A momentum space with the topology of a four-sphere and radius $\propto\k$ has a two-sphere boundary momentum space and $\zeta_2(s)\propto\k^{3-2s}\Gamma(1-s)/[\Gamma(3/2-s)]$. The pole $s=1$ corresponds to $\ds^{\rm b}=2$, the spectral dimension of the boundary in the IR. In the UV $\ds^{\rm b}\simeq 0$, $\zeta_2(0)$ is finite, while $\rho_2$ can be shown to be ill defined, despite the theory being UV finite \cite{ArCa3}. Thus, models with compact momentum space on the boundary exhibit a vanishing spectral dimension for small diffusion times, which leads to a divergent entanglement entropy density.

The finite-$\rho_d$ condition we just found is necessary but not sufficient. Indeed, even when the spectral dimension does not vanish at any scale, it may be possible that the entanglement entropy density is divergent, as in ordinary local quantum field theory (constant $\ds\neq 0$).

It seems that having a running spectral dimension is crucial if we want to get a finite entropy density. It is no wonder, then, that basically all known quantum-gravity models exhibit this feature (see questions {\it 03} and {\it 48} of \cite{revmu} and references therein). On one hand, a finite $\rho_d$ is conjectured to signal a quantization of gravitational degrees of freedom and, on the other hand, in various approaches to quantum gravity the spectral dimension of spacetime always runs according to the resolution scale.

\begin{figure}
\centering
\includegraphics[width=10cm]{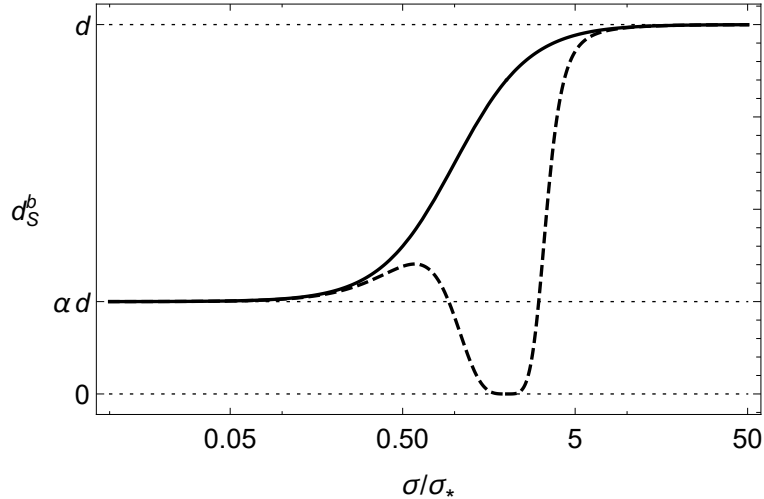}
\caption{\label{fig1} The typical kink profile of the boundary spectral dimension in many quantum gravities (solid curve). This is reproduced by the multi-fractional theory with $q$-derivatives, whose zeta function $\zeta_d(s)\propto\Gamma\left[\frac{d-2s}{2(1-\a)}\right]\Gamma\left[\frac{2s-d\a}{2(1-\a)}\right]/\Gamma(s)$ has poles at $s=d/2=\ds^{\rm b,IR}/2$ and $s=d\a/2=\ds^{\rm b,UV}/2$, where $0<\a<1$. The dashed curve represents a profile of the spectral dimension vanishing at some intermediate scale, leading to an infinite entanglement entropy density.}
\end{figure}

It is instructive to consider a theory respecting our necessary condition and producing a finite entanglement entropy, but which however is not power-counting renormalizable. We are talking about the so-called multi-fractional theory with $q$-derivatives. This framework, reviewed in \cite{revmu}, is a field theory defined on a multi-fractal geometry with varying Hausdorff and spectral dimension. Dimensional flow follows a multi-parametric universal profile determined only by very general (background- and dynamics-independent) properties of the spacetime dimension. Spacetime is characterized by at least one fundamental length scale $\ell_*\propto\sqrt{\s_*}$, above which ordinary geometry is recovered, while at scales $\lesssim\ell_*$ the spacetime dimensionality drops below $D$. At ultra-microscopic scales, a discrete structure naturally emerges, encoded in logarithmic oscillations of the geometry. Physical observables are affected by this scale dependence. The generality of the measure, together with the fact that all quantum-gravity models exhibit dimensional flow, justifies the interest in these theories as an efficient framework wherein to explore the physical consequences of dimensional flow and to constrain them with experiments and observations ranging from particle physics to cosmology. After an analytic continuation, the entanglement entropy density of this theory for an isotropic spacetime measure is \cite{ArCa3}
\be\label{rhoda}
\rho_d\propto \Gamma\left[-\frac{d\a}{2(1-\a)}\right]\,,
\ee
where $0<\a<1$ parametrizes the spectral dimension $\ds^{\rm b}\simeq d\a$ of the boundary in the UV (in the IR, $\ds^{\rm b}\simeq d$; Fig.\ \ref{fig1}). When $\a=[1+d/(2n)]^{-1}$ (which includes the special case $n=0$, $\a=0$), this expression diverges, but it is finite otherwise. Since gravity cannot be easily quantized perturbatively in this scenario \cite{revmu}, this example shows that imposing a finite entanglement entropy density does not imply the existence of a well defined quantization of the gravitational degrees of freedom. 

Our arguments show that the requirement of a finite entanglement entropy density is quite a restrictive one and demanding UV finiteness is certainly not enough. Even if the example of fractals \cite{Akk12} indicates that, at least in principle, a varying spectral dimension it is not indispensable if the underlying geometry is discrete, in physical theories the role of dimensional flow seems to be that of a key actor. Also, the balance between the level of smoothness of geometry and the way geometry changes with the probed scale is delicate and, ultimately, it is relevant for the emergence of spacetime (of which locality and smoothness are key aspects \cite{HaMa}). The entanglement entropy itself governs such emergence quantitatively, prescribing how two spacetime regions can connect or are tore away classically depending on whether the degrees of freedom within these regions entangle and disentangle at the quantum level \cite{Van10}. Since dimensional flow is a potentially \emph{observable} feature of quantum gravity models \cite{revmu,CaRo2a}, the interest in its relation with the entanglement entropy is not merely academic and it could have a significant impact on the search of experimental signatures of quantum gravity. The present result, further extended in \cite{ArCa3}, aims to set the stage for near-future studies on this fascinating topic.

\bigskip

\paragraph*{Acknowledgments.} The work of G.C.\ is under a Ram\'on y Cajal contract and is supported by the I+D grant FIS2014-54800-C2-2-P.


\begin{thebibliography}{99}
\bibitem{BH1} J.D.\ Bekenstein, \tia{Black holes and entropy} \doin{10.1103/PhysRevD.7.2333}{Phys.\ Rev.}{D}{7}{2333}{1973}.
\bibitem{BH2} S.W.\ Hawking, \tia{Particle creation by black holes} \doinn{10.1007/BF02345020}{Commun.\ Math.\ Phys.}{43}{199}{1975}; \doinn{10.1007/BF01608497}{}{46}{206}{1976}.
\bibitem{StVa}  A.\ Strominger and C.\ Vafa, \tia{Microscopic origin of the Bekenstein--Hawking entropy} \doin{10.1016/0370-2693(96)00345-0}{Phys.\ Lett.}{B}{379}{99}{1996} [\oarX{hep-th/9601029}].
\bibitem{ABCK}  A.\ Ashtekar, J.C.\ Baez, A.\ Corichi and K.\ Krasnov, \tia{Quantum geometry and black hole entropy} \doinn{10.1103/PhysRevLett.80.904}{Phys.\ Rev.\ Lett.}{80}{904}{1998} [\oarX{gr-qc/9710007}].
\bibitem{OrPS}  D.\ Oriti, D.\ Pranzetti and L.\ Sindoni, \tia{Horizon entropy from quantum gravity condensates} \doinn{10.1103/PhysRevLett.116.211301}{Phys.\ Rev.\ Lett.}{116}{211301}{2016} [\arX{1510.06991}].
\bibitem{CHRR}  G.\ Chirco, H.M.\ Haggard, A.\ Riello and C.\ Rovelli, \tia{Spacetime thermodynamics without hidden degrees of freedom} \doin{10.1103/PhysRevD.90.044044}{Phys.\ Rev.}{D}{90}{044044}{2014} [\arX{1401.5262}].
\bibitem{CR}    S.M.\ Carroll and G.N.\ Remmen, \tia{What is the entropy in entropic gravity?} \doin{10.1103/PhysRevD.93.124052}{Phys.\ Rev.}{D}{93}{124052}{2016} [\arX{1601.07558}].
\bibitem{Casini:2009sr} H.\ Casini and M.\ Huerta, \tia{Entanglement entropy in free quantum field theory} \doin{10.1088/1751-8113/42/50/504007}{J.\ Phys.}{A}{42}{504007}{2009} [\arX{0905.2562}].
\bibitem{Solodukhin:2011gn} S.N.\ Solodukhin, \tia{Entanglement entropy of black holes} \doinn{10.12942/lrr-2011-8}{Living Rev.\ Rel.}{14}{8}{2011} [\arX{1104.3712}].
\bibitem{Frolov:1998ea} V.P.\ Frolov and D.\ Fursaev, \tia{Black hole entropy in induced gravity: reduction to 2D quantum field theory on the horizon} \doin{10.1103/PhysRevD.58.124009}{Phys.\ Rev.}{D}{58}{124009}{1998} [\oarX{hep-th/9806078}].
\bibitem{Nesterov:2010jh} D.\ Nesterov and S.N.\ Solodukhin, \tia{Short-distance regularity of Green's function and UV divergences in entanglement entropy} \doij{10.1007/JHEP09(2010)041}{JHEP}{1009}{041}{2010} [\arX{1008.0777}].
\bibitem{KowalskiGlikman:2004tz} J.\ Kowalski-Glikman and S.\ Nowak, \tia{Quantum $\kappa$-Poincar\'e algebra from de Sitter space of momenta} \oarX{hep-th/0411154}.
\bibitem{Girelli:2009yz} F.\ Girelli, E.R.\ Livine and D.\ Oriti, \tia{Four-dimensional deformed special relativity from group field theories} \doin{10.1103/PhysRevD.81.024015}{Phys.\ Rev.}{D}{81}{024015}{2010} [\arX{0903.3475}].
\bibitem{Arzano:2010jw} M.\ Arzano, \tia{Anatomy of a deformed symmetry: field quantization on curved momentum space} \doin{10.1103/PhysRevD.83.025025}{Phys.\ Rev.}{D}{83}{025025}{2011} [\arX{1009.1097}].
\bibitem{Alesci:2011cg} E.\ Alesci and M.\ Arzano, \tia{Anomalous dimension in semiclassical gravity} \doin{10.1016/j.physletb.2011.12.026}{Phys.\ Lett.}{B}{707}{272}{2012} [\arX{1108.1507}].
\bibitem{Arzano:2014jfa} M.\ Arzano and T.\ Trze\'sniewski, \tia{Diffusion on $\kappa$-Minkowski space} \doin{10.1103/PhysRevD.89.124024}{Phys.\ Rev.}{D}{89}{124024}{2014} [\arX{1404.4762}].
\bibitem{Amelino-Camelia:2013gna} G.\ Amelino-Camelia, M.\ Arzano, G.\ Gubitosi and J.\ Magueijo, \tia{Dimensional reduction in momentum space and scale invariant cosmological fluctuations} \doin{10.1103/PhysRevD.88.103524}{Phys.\ Rev.}{D}{88}{103524}{2013} [\arX{arXiv:1309.3999}].
\bibitem{ArCa3} M.\ Arzano and G.\ Calcagni, \tia{Finite entanglement entropy and spectral dimension in quantum gravity} \arX{1704.01141}.
\bibitem{tH93}  G.\ 't Hooft, \tia{Dimensional reduction in quantum gravity} \procsinm{Salamfestschrift}{A.\ Ali, J.\ Ellis, S.\ Randjbar-Daemi}{World Scientific}{Singapore}{1993} [\oarX{gr-qc/9310026}].
\bibitem{Car09} S.\ Carlip, \tia{Spontaneous dimensional reduction in short-distance quantum gravity?} \doinn{10.1063/1.3284402}{AIP Conf.\ Proc.}{1196}{72}{2009} [\arX{0909.3329}].
\bibitem{revmu} G.\ Calcagni, \tia{Multifractional theories: an unconventional review} \doij{10.1007/JHEP03(2017)138}{JHEP}{1703}{138}{2017} [\arX{1612.05632}].
\bibitem{Akk12} E.\ Akkermans, \tia{\href{http://www.ams.org/books/conm/601/11962/conm601-11962.pdf}{\cob Statistical mechanics and quantum fields on fractals}} \procm{Fractal Geometry and Dynamical Systems in Pure and Applied Mathematics II: Fractals in Applied Mathematics}{D.\ Carfi, M.L.\ Lapidus, E.P.J.\ Pearse and M.\ van Frankenhuijsen}{AMS}{Providence}{U.S.A.}{2013} [\arX{1210.6763}].
\bibitem{HaMa}  S.A.\ Hartnoll and E.\ Mazenc, \tia{Entanglement entropy in two-dimensional string theory} \doinn{10.1103/PhysRevLett.115.121602}{Phys.\ Rev.\ Lett.}{115}{121602}{2015} [\arX{1504.07985}].
\bibitem{Van10} M.\ Van Raamsdonk, \tia{Building up spacetime with quantum entanglement} \doinn{10.1007/s10714-010-1034-0}{Gen.\ Rel.\ Grav.}{42}{2323}{2010} [\doin{10.1142/S0218271810018529}{Int.\ J.\ Mod.\ Phys.}{D}{19}{2429}{2010}] [\arX{1005.3035}].
\bibitem{CaRo2a} G.\ Amelino-Camelia, G.\ Calcagni and M.\ Ronco, \tia{Imprint of quantum gravity in the dimension and fabric of spacetime} \arX{1705.04876}.
\end{thebibliography}
\end{document}